\documentclass[11pt, oneside]{article}   	
\usepackage{geometry}                		
\usepackage{graphicx}				
\usepackage{amssymb}      
\usepackage{amsmath}
\usepackage{empheq}
\usepackage{amsthm}
\usepackage[utf8]{inputenc}
\usepackage[final]{pdfpages}
\newtheorem{theorem}{Theorem}
\newtheorem*{theorem*}{Theorem}

\begin{document}
\title{On a class of integrable Hamiltonian equations in 2+1 dimensions }

\author{B. Gormley$^{1}$, E.V. Ferapontov$^{1, 2}$, V.S. Novikov$^{1}$
}
     \date{}
     \maketitle
     \vspace{-5mm}
\begin{center}
$^1$Department of Mathematical Sciences \\ Loughborough University \\
Loughborough, Leicestershire LE11 3TU \\ United Kingdom \\
$^2$Institute of Mathematics, Ufa Federal Research Centre,\\
Russian Academy of Sciences, 112, Chernyshevsky Street, \\ Ufa 450077, Russia \vspace{5pt}\\

e-mails: \\[1ex]
\texttt{B.Gormley@lboro.ac.uk}\\
\texttt{E.V.Ferapontov@lboro.ac.uk}\\
\texttt{V.S.Novikov@lboro.ac.uk}\\
\end{center}

\bigskip

\begin{abstract}
We classify integrable Hamiltonian equations of the form
$$
u_t=\partial_x\left(\frac{\delta H}{\delta u}\right), \quad H= \int h(u, w)\ dxdy,
$$
where the Hamiltonian density $h(u, w)$ is a function of two variables: dependent variable $u$ and  the non-locality $w=\partial_x^{-1}\partial_yu$. Based on the method of hydrodynamic reductions, the integrability conditions  are derived (in the form of an involutive PDE system for the  Hamiltonian density $h$). We show that the generic integrable density is expressed in terms of the Weierstrass $\sigma$-function: $h(u, w)=\sigma(u)e^w$. Dispersionless Lax pairs, commuting flows and dispersive deformations of the resulting equations are also discussed.

\bigskip


\bigskip

\end{abstract}

\newpage

\centerline{\it To Allan Fordy on the occasion of his 70th birthday}

\tableofcontents

\section{Introduction}

 \subsection{Formulation of the problem}
In this paper we investigate Hamiltonian systems of the form 
 \begin{equation}
 u_t = \partial_x\bigg(\frac{\delta H}{\delta u}\bigg), \quad  H = \int h(u,w)\ dxdy. \label{eq:pde1}
 \end{equation}
Here $\partial_x$ is the Hamiltonian operator, and the Hamiltonian density $h(u, w)$ depends on $u$ and  the nonlocal variable $w=\partial_x^{-1}\partial_yu$    (equivalently, $w_x = u_y$). Since 
$
\frac{\delta H}{\delta u} =h_u +\partial_x^{-1}\partial_y(h_w) 
$
we can rewrite equation (\ref{eq:pde1}) in the two-component first-order quasilinear form:
 \begin{align}
 u_t = (h_u)_x +(h_w)_y, \quad w_x=u_y.   \label{eq:pde2} 
 \end{align}
Familiar examples of this type include the dispersionless KP equation ($h=\frac{1}{2}w^2+\frac{1}{6}u^3$) and the dispersionless Toda (Boyer-Finley) equation ($h=e^w$). Our main goal  is to classify integrable systems within class (\ref{eq:pde2}) and to construct their dispersionless Lax pairs, commuting flows and dispersive deformations. Before stating our main results, let us begin with a brief description of the existing approaches to dispersionless integrability in 2+1 dimensions.

\subsection{Equivalent approaches to dispersionless integrability} 

Here we summarise three existing approaches to integrability of equations of type (\ref{eq:pde2}), namely, the method of hydrodynamic reductions, the geometric approach based on integrable conformal  geometry (Einstein-Weyl geometry), and the method of dispersionless Lax pairs. Based on seemingly different ideas, these approaches  lead to equivalent integrability conditions/classification results \cite{FK}.

\medskip

\noindent{\bf The method of hydrodynamic reductions}, see e.g. \cite{GT, Fer1}, consists of seeking multiphase solutions to system (\ref{eq:pde2}) in the form 
\begin{equation}\label{uw}
u=u(R^1,R^2,\ldots,R^n), \quad w=w(R^1,R^2,\ldots,R^n)
\end{equation}
where the phases $R^i(x,y,t)$ (also known as Riemann invariants; note that their number $n$ can be arbitrary) satisfy a pair of commuting hydrodynamic-type systems: 
\begin{equation}\label{R}
R^i_y = \mu^i(R)R^i_x, \quad
R^i_t = \lambda^i(R)R^i_x; 
\end{equation}
we recall that the commutativity conditions are equivalent to the following constraints for the characteristic speeds $\mu^i, \lambda^i$ \cite{T, T1}:
\begin{equation}
\frac{\partial_j\mu^i}{\mu^j - \mu^i} = \frac{\partial_j\lambda^i}{\lambda^j - \lambda^i}, \label{eq:hydroredu3} 
\end{equation}
$i\ne j, \ \partial_j=\partial_{R^j}$.  Substituting ansatz (\ref{uw}) into  (\ref{eq:pde2}) and using (\ref{R}), (\ref{eq:hydroredu3}) one obtains an overdetermined system  for  the unknowns $u, w, \mu^i, \lambda^i$, viewed as functions of $R^1, \dots, R^n$ (the so-called generalised Gibbons-Tsarev system, or GT-system). 
System (\ref{eq:pde2}) is said to be integrable by the method of hydrodynamic reductions if it possesses `sufficiently many' multi-phase solutions of  type (\ref{uw}), in other words, if the corresponding GT-system is involutive. Note that the coefficients of GT-system will depend on the density $h(u, w)$ and its partial derivatives. The requirement
that GT-system  is involutive imposes differential constraints for the Hamiltonian density $h$, the so-called integrability conditions. Details of this computation will be given in Section \ref{sec:hydro}.

\medskip

\noindent{\bf Integrability via Einstein-Weyl geometry.}  Let us first introduce a conformal structure defined by the characteristic variety of system (\ref{eq:pde2}). Given a  $2\times 2$ quasilinear system 
\begin{equation*}
A(v)v_{x^1} + B(v)v_{x^2} + C(v)v_{x^3} = 0 
\end{equation*} 
where $A,B,C$  are $2\times 2$ matrices  depending on $v=(u, w)^T$, the characteristic equation of this system, $\det(Ap_1 + Bp_2 + Cp_3) = 0$, defines a conic $g^{ij}p_ip_j =0$. This gives the  characteristic conformal structure  $[g] = g_{ij}dx^idx^j$ where $g_{ij}$ is the inverse matrix of $g^{ij}$. For system (\ref{eq:pde2}) direct calculation gives 
\begin{equation}\label{[g]}
[g] = 4h_{ww}dxdt - dy^2 - 4h_{uw}dydt + 4(h_{ww}h_{uu} - h_{uw}^2)dt^2;  
\end{equation} 
here we set $(x^1, x^2, x^3)=(x, y, t)$. Note that $[g]$ depends upon a solution to the system (we will assume $[g]$ to be non-degenerate, which is equivalent to the condition $h_{ww}\ne 0$). It turns out that  integrability of system (\ref{eq:pde2}) can be reformulated geometrically as the Einstein-Weyl property of the characteristic conformal structure $[g]$.  We recall that Einstein-Weyl geometry is a triple $(\mathbb{D},[g],\omega)$ where $[g]$ is a conformal structure, $\mathbb{D}$ is a symmetric  affine connection  and $\omega = \omega_k dx^k$ is a 1-form  such that 
\begin{equation}\label{EW}
\mathbb{D}_kg_{ij} = \omega_kg_{ij}, \quad
R_{(ij)} = \Lambda g_{ij} 
\end{equation}
for some function $\Lambda$ \cite{Cartan, Cartan1}; here $R_{(ij)}$ is the symmetrised Ricci tensor of $\mathbb{D}$. Note that the first part of equations (\ref{EW}), known as the Weyl equations, uniquely determines $\mathbb{D}$ once $[g]$ and $\omega$ are specified.  It was observed in \cite{FK} that for broad classes of dispersionless integrable systems (in particular, for systems of type (\ref{eq:pde2})), the one-form  $\omega$ is given in terms of $[g]$ by a universal explicit formula
\begin{equation*}
\omega_k = 2g_{kj}\mathcal{D}_s g^{js} + \mathcal{D}_k \ln{(\det g_{ij})}
\end{equation*}   
where $\mathcal{D}_k$ denotes the total derivative with respect to $x^k$. Applied to $[g]$ given by (\ref{[g]}), this formula implies
\begin{align}
\omega_1&=0, \nonumber \\ 
\omega_2&=\frac{2(h_{uuw}v_{xx} + 2h_{uww} v_{xy} + h_{www} v_{yy})}{h_{ww}}, \label{om} \\
\omega_3&=\frac{4(h_{uw}(h_{uuw}v_{xx} + 2h_{uww}v_{xy} + h_{www}v_{yy})}{h_{ww}} - \frac{ h_{ww}(h_{www}v_{xx} + 2h_{uuw}v_{xy} + h_{uww} v_{yy} ))}{h_{ww}}. \nonumber
\end{align}
 To summarise,  integrability of system (\ref{eq:pde2}) is equivalent to the Einstein-Weyl property of $[g],\ \omega$ given by (\ref{[g]}), (\ref{om}) on every solution of system (\ref{eq:pde2}). Note that in 3D, Einstein-Weyl equations (\ref{EW}) are themselves integrable by the twistor construction \cite{Hitchin}, see also \cite{DFK}, and thus constitute `integrable conformal geometry'.

\medskip

\noindent{\bf Dispersionless Lax pair}  of system  (\ref{eq:pde2}) consist of two Hamilton-Jacobi type equations for an auxiliary function $S$, 
$$
S_t = F(S_x,u,w), \quad S_y = G(S_x,u,w),
$$
whose compatibility condition, $S_{ty} = S_{yt}$, is equivalent to system (\ref{eq:pde2}). Dispersionless Lax pairs  were introduced in \cite{Zakharov} as quasiclassical limits of Lax pairs of integrable soliton equations in 2+1D. It is known that the existence of a dispersionless Lax representation is equivalent to
hydrodynamic/geometric integrability discussed above \cite{Fer2, FK}.  
We refer to Section \ref{sec:disp} for dispersionless Lax pairs of integrable systems (\ref{eq:pde2}).

\subsection{Summary of the main results}

Our first result is the  set of integrability conditions for the Hamiltonian density $h$.

\begin{theorem} \label{t1} The following conditions are equivalent:

\noindent (a) System (\ref{eq:pde2}) is integrable by the method of hydrodynamic reductions;

\noindent (b) Characteristic conformal structure [g] and covector $\omega$ given by (\ref{[g]}), (\ref{om}) satisfy Einstein-Weyl equations (\ref{EW}) on every solution of system  (\ref{eq:pde2}); 

\noindent (c) System (\ref{eq:pde2}) possesses a dispersionless Lax pair;

\noindent (d) Hamiltonian density $h(u, w)$ satisfies the following set of integrability conditions: 
\begin{align}
h_{www}^2 - h_{ww}h_{wwww} &= 0, \nonumber \\ 
h_{uww}h_{www} - h_{ww}h_{uwww} &= 0, \nonumber \\
h_{uuw}h_{www} - h_{ww}h_{uuww} &= 0, \label{int} \\
h_{uuu}h_{www} - h_{ww}h_{uuuw} &=0, \nonumber \\
-3 h_{uuw}^2 + 4 h_{uww}h_{uuu} -h_{ww}h_{uuuu} &= 0. \nonumber
\end{align}
\end{theorem}
Theorem \ref{t1} is proved in Section \ref{sec:hydro}. The system of integrability conditions (\ref{int}) is involutive, and modulo natural equivalence transformations its solutions can be reduced to one of the six canonical forms.

\begin{theorem} \label{t2} Solutions  $h(u, w)$ of system (\ref{int}) can be reduced to  one of the six canonical forms: 
\begin{align*}
h(u,w) &= \frac{1}{2}w^2 + \frac{1}{6}u^3,\\
h(u,w) &= w^2 + u^2 w - \frac{1}{4}u^4, \\
h(u,w) &= uw^2  + \beta u^7, \\
h(u,w) &= e^w,\\
h(u,w) &= ue^w, \\
h(u,w) &= \sigma(u;0,g_3)e^w;
\end{align*}
here $\beta$ and $g_3$  are constants, and $\sigma(u; g_2, g_3)$ denotes the Weierstrass sigma function. 
\end{theorem}
Theorem \ref{t2} is proved in Section \ref{sec:can}. Dispersionless Lax pairs for the corresponding systems (\ref{eq:pde2}) are constructed in Section \ref{sec:disp}. 

It turns out that every integrable system (\ref{eq:pde2}) possesses a higher commuting  flow of the form
\begin{align}
u_{\tau} &= a(u,w,v)u_x + b(u,w,v)u_y + c(u,w,v)w_y + d(u,w,v)v_y, \nonumber \\  
w_x &= u_y, \label{com}\\
v_x &= (p(u,w))_y, \nonumber
\end{align}
where $\tau$ is the higher `time', and $v=\partial_x^{-1}\partial_y p(u, w)$ is an extra nonlocal variable
(in contrast to the 1+1 dimensional case, higher commuting flows in 2+1 dimensions require higher nonlocalities). Remarkably, 
the structure of higher nonlocalities is uniquely determined by the original system (\ref{eq:pde2}), in particular, the function $p(u, w)$ can be expressed in terms of $h(u, w)$: $p=h_w$. Furthermore, commuting flow
(\ref{com}) is automatically Hamiltonian. 

\begin{theorem} \label{t3} 
Every integrable  system (\ref{eq:pde2}) possesses a higher commuting flow (\ref{com}) with the nonlocality $v_x = (h_w)_y$. Commuting flow (\ref{com}) is Hamiltonian  with the Hamiltonian density $f(u,w,v)$ of the form
\begin{equation*}
f(u,w,v)= vh_w(u,w)  + g(u,w),
\end{equation*}
where $g(u,w)$ can be recovered from the compatible equations
\begin{align*}
g_{ww} &= 4h_{uw}h_{ww} + \alpha wh_{ww}, \\
g_{uuu} &= 8h_{uw}h_{uuu} + \alpha wh_{uuu}, \\
g_{uuw} &= 6h_{uw}h_{uuw} + 2h_{ww}h_{uuu} + \alpha w h_{uuw}.
\end{align*}
Here the constant $\alpha$ is defined by the relation $\alpha = 2h_{ww}\frac{\partial}{\partial w}\big(\frac{h_{uw}}{h_{ww}}\big)$ which follows from integrability conditions (\ref{int}).

\end{theorem} 
Theorem \ref{t3} is proved in Section \ref{sec:comm}. Dispersionless Lax pairs for commuting flows are constructed in Section \ref{sec:dispcomm}.

\section{Proofs}

In this section we prove Theorems \ref{t1}-\ref{t3} and construct Lax pairs for integrable systems (\ref{eq:pde2}) and their commuting flows (\ref{com}).

\subsection{The method of hydrodynamic reductions: proof of Theorem \ref{t1}}
\label{sec:hydro}

\noindent {\bf Equivalences (a) $\Leftrightarrow$ (b) and (a) $\Leftrightarrow$ (c)} of Theorem \ref{t1} follow from the  results of \cite{FK} and \cite{Fer2} which hold for general two-component systems of hydrodynamic type in 2+1 dimensions. 

\medskip

\noindent {\bf Equivalence (a) $\Leftrightarrow$ (d)} can be demonstrated as follows. Let us rewrite  system (\ref{eq:pde2}) in the form 
$$
u_t = h_{uu}u_x + 2h_{uw}u_y + h_{ww}w_y,\quad
w_x = u_y,
$$
and substitute the ansatz $u=u(R^1,R^2,\ldots,R^n), \ w=w(R^1,R^2,\ldots,R^n)$. Using equations (\ref{R}) and collecting coefficients at $R^i_x$ we obtain  $\partial_iw= \mu^i\partial_iu$, along with the dispersion relation $\lambda^i = h_{uu} + 2h_{uw}\mu^i + h_{ww}(\mu^i)^2$. Substituting the last formula  into the commutativity conditions  (\ref{eq:hydroredu3}) we obtain 
\begin{equation}\label{GT1}
\footnotesize{\partial_j\mu^i = \frac{h_{uuu} + h_{uuw}(\mu^j + 2\mu^i) + h_{uww}\big(2\mu^i\mu^j + (\mu^i)^2\big) + h_{www}\mu^j(\mu^i)^2}{h_{ww}(\mu^j -\mu^i)}\partial_ju. }
\end{equation} 
Finally, the compatibility condition $\partial_i\partial_jw = \partial_j\partial_iw$ results in 
\begin{equation}\label{GT2}
\footnotesize{\partial_i\partial_ju =\frac{2h_{uuu} + 3h_{uuw}(\mu^j + \mu^i) + h_{uww}((\mu^i)^2 + 4\mu^i\mu^j + (\mu^j)^2) + h_{www}(\mu^j(\mu^i)^2 + \mu^i(\mu^j)^2)}{h_{ww}(\mu^j -\mu^i)^2}\partial_iu\partial_ju}.
\end{equation}
Equations (\ref{GT1}), (\ref{GT2}) constitute the corresponding GT-system. As one can see, it contains partial derivatives of the Hamiltonian density $h$ in the coefficients. Verifying involutivity of GT-system amounts to 
checking the compatibility conditions $\partial_k(\partial_j\mu^i) =\partial_j(\partial_k\mu^i)$ and $\partial_k(\partial_i\partial_ju) =\partial_j(\partial_i\partial_ku)$. Direct computation (performed in Mathematica)  results in the integrability conditions (\ref{int}) for $h(u, w)$.  Note that without any loss of generality one can restrict to the case when the number of Riemann invariants $R^i$ is equal to three, indeed, all compatibility conditions involve three distinct indices only. This finishes the proof of Theorem \ref{t1}.

\subsection{Canonical forms of integrable  densities: proof of Theorem \ref{t2}}
\label{sec:can}
We have five integrability conditions, namely 
\begin{align}
h_{www}^2 - h_{ww}h_{wwww} &= 0, \label{eq:int1} \\
h_{uww}h_{www} - h_{ww}h_{uwww} &= 0, \label{eq:int2}\\ 
h_{uuw}h_{www} - h_{ww}h_{uuww} &= 0, \label{eq:int3} \\
h_{uuu}h_{www} - h_{ww}h_{uuuw} &=0, \label{eq:int4}\\
-3h_{uuw}^2 + 4 h_{uww}h_{uuu} -h_{ww}h_{uuuu} \label{eq:int5} &= 0. 
\end{align}
The classification of solutions will be performed modulo equivalence transformations leaving system (\ref{eq:pde2}) form-invariant (and therefore preserving the integrability conditions). These include 
\begin{equation}\label{tr1}
\tilde x=x-2a t, \quad \tilde y=y-2b t, \quad \tilde h= h+a u^2+b uw+m u+n w+p,
\end{equation}
as well as
\begin{equation}\label{tr2}
\tilde x=x-s y, \quad \tilde w=w+s u; 
\end{equation}
(other variables remain unchanged). We will always assume $h_{ww}\ne 0$ which is equivalent to the requirement of irreducibility of the dispersion relation. There are two main cases to consider.

\medskip 
\noindent{\bf Case 1: $h_{www} =0.$}  Then
\begin{equation*}
h(u,w) = \alpha(u)w^2 + \beta(u)w + \gamma(u), 
\end{equation*}
and the integrability conditions imply 
$$
\alpha'' = 0, \quad \beta'''= 0, \quad -3\beta''^2+8\alpha'\gamma'''-2\alpha \gamma ''''=0.
$$
There are two further subcases: $\alpha =1$ and $\alpha=u$. 

The subcase $\alpha =1$ leads, modulo equivalence transformations (\ref{tr1}), to densities of the form
\begin{equation*}
h(u,w) = w^2 + \beta_1 u^2w  - \frac{\beta_1^2}{4}u^4 + \gamma_1 u^3, 
\end{equation*}
$\beta_1, \gamma_1 = const$.  For $\beta_1 =0$ we obtain the first case of Theorem \ref{t2} (after a suitable rescaling). If $\beta_1 \ne 0$ then we can eliminate the term $u^3$ by a translation of $u$. This gives the second case of Theorem \ref{t2} (after rescaling of $u$ and $w$).

The subcase $\alpha =u$ leads, modulo equivalence transformations (\ref{tr1}), to densities of the form
\begin{equation*}
h(u,w) = uw^2 + \beta_1 u^2w + \gamma_1u^7 + \frac{\beta_1^2}{4}u^3.
\end{equation*}
$\beta_1, \gamma_1 = const$. Note that we can set $\beta_1=0$ using transformation (\ref{tr2}) with $s=\beta_1/2$. This gives the third case of Theorem \ref{t2}.

\medskip 
\noindent{\bf Case 2: $h_{www} \neq 0.$} Then the first two integrability conditions  $(\ref{eq:int1})$ and $(\ref{eq:int2})$  imply $h_{www} = c h_{ww}$ for some constant $c$ (which can be set equal to $1$). This gives 
$$
h(u,w) = a(u)e^{w} +p(u)w + q(u).
$$
The next two integrability conditions (\ref{eq:int3}) and (\ref{eq:int4}) give $p'' = 0$ and $q''' = 0$, respectively. Thus, modulo equivalence transformations (\ref{tr1}) we can assume $h(u, w)=a(u)e^w$.
Finally, equation (\ref{eq:int5}) implies
\begin{equation*}
aa'''' - 4 a'a''' + 3 a''^2 = 0,
\end{equation*}
which is the classical equation for the Weierstrass sigma function (equianharmonic case $g_2=0$). 
Setting $\wp = - (\ln a)''$ we obtain $\wp'' = 6 \wp^2$, which integrates to 
\begin{equation}
\wp'^2  = 4 \wp^3 - g_3,
\end{equation}
$g_3=const$. 
There are three subcases. 
\medskip

\noindent {\it Subcase $ g_3 =0,\ \wp = 0$.}
Then $a(u) = e^{\alpha u + \beta}$ and modulo equivalence transformations (\ref{tr2}) we obtain Case 4 of Theorem \ref{t2}.  

\medskip

\noindent{\it Subcase $ g_3 =0,\ \wp = \frac{1}{u^2}$.} Then $a(u) = ue^{\alpha u + \beta}$ and modulo equivalence transformations (\ref{tr2}) we obtain Case 5 of Theorem \ref{t2}. 

\medskip

\noindent{\it Subcase $ g_3 \neq 0$.}
Then $a(u) = \sigma(u;0,g_3)e^{\alpha u + \beta}$ and modulo equivalence transformations (\ref{tr2}) we obtain the last  case of our classification. This finishes the proof of Theorem \ref{t2}.

\medskip

\noindent{\bf Remark.} The paper \cite{FOS} gives a classification of integrable two-component Hamiltonian systems of the form 
\begin{equation}\label{h2}
\renewcommand\arraystretch{1.4}
\begin{bmatrix}
U_t \\
W_t
\end{bmatrix}
= 
\begin{bmatrix}
0 & \partial_x \\
\partial_x & \partial_y
\end{bmatrix}
\begin{bmatrix}
\frac{\delta H}{\delta U} \\ 
\frac{\delta H}{\delta W} 
\end{bmatrix}
\end{equation}
where $H=\int F(U, W)\ dxdy$. Explicitly, we have
$$
U_t = (F_W)_x, \quad
W_t = (F_U)_x + (F_W)_y. 
$$
Let us introduce a contact change of variables $(U, W, F)\to (u, w, f)$ via partial  Legendre transform:
$$
w = F_W, \quad u = U, \quad f= F - WF_W, \quad f_w = -W, \quad f_u = F_U. 
$$
In the new variables the system becomes 
$$
w_y = -(f_u)_x - (f_w)_t, \quad u_t = w_x. 
$$
Modulo relabelling $u\leftrightarrow w,\ f\to -h,\ y\to t, \ t\to x, \ x \to y$ these equations coincide with (\ref{eq:pde2}). Thus, Hamiltonian formalisms (\ref{eq:pde1}) and (\ref{h2}) are equivalent. Examples of  dKP and Boyer-Finley equations suggest however that Hamiltonian formalism (\ref{eq:pde1}) is more natural and convenient, indeed, in the form (\ref{eq:pde1}) both equations  arise directly in their `physical' variables.

\subsection{Dispersionless Lax pairs} 
\label{sec:disp}
In this section we provide dispersionless Lax representations for all six canonical forms of 
Theorem \ref{t2}. The results are summarised in  Table 1 below. 

\medskip

 \centerline{\footnotesize{Table 1: Dispersionless Lax pairs for integrable systems (\ref{eq:pde2})}}
 
\begin{center}
    \begin{tabular}{  | p{6.4cm} | p{7.4cm} |}
    \hline
     Hamiltonian density $h(u, w)$  & Dispersionless Lax pair  \\ \hline
       $h(u,w) = \frac{1}{2}w^2 + \frac{1}{6}u^3$ & \\ 
     ${\rm System}\ (\ref{eq:pde2}):$ & $S_t = \frac{1}{3}{S_x^3} + u S_x + w$ \\
    $u_t=uu_x+w_y$ & $S_y = \frac{1}{2}S_x^2 +u$\\
    $w_x=u_y$ &   \\ 
      \hline 
   $h(u,w) = w^2 +  u^2 w - \frac{1}{4}u^4$ & \\ 
 ${\rm System}\ (\ref{eq:pde2}):$ & $S_t = (3 u^2 + 2  w)S_x + 2  u S_x^4 +\frac{2}{7}S_x^7$ \\
    $u_t=(2 w-3u^2)u_x+4 u u_y+2w_y$ & $S_y =  u S_x + \frac{1}{4}S_x^4$\\
    $w_x=u_y$ &   \\ 
      \hline 
       $h(u,w) = uw^2  + \beta u^7$ &  \\ 
  ${\rm System}\ (\ref{eq:pde2}):$ & $S_t = 4u^2\wp(S_x)(w + \frac{1}{5}u^3\wp'(S_x))$ \\
    $u_t=42\beta u^5u_x+4wu_y+2uw_y$ & $S_y = u^2 \wp(S_x)$\\
    $w_x=u_y$ &  here $\wp'^2=4\wp^3-35\beta$ \\ 
      \hline 
             $h(u,w) = e^w$ &  \\ 
  ${\rm System}\ (\ref{eq:pde2}):$ & $S_t = -\frac{e^w}{S_x + u} $ \\
    $u_t=e^ww_y$ & $S_y =  -\ln(S_x + u)$\\
    $w_x=u_y$ &   \\ 
      \hline        $h(u,w) = ue^w$ &  \\ 
 ${\rm System}\ (\ref{eq:pde2}):$ & $S_t = \frac{ 3u^2 e^w S_x}{u^3 - S_x^3}$ \\
    $u_t=e^w(2u_y+uw_y)$ & $S_y =\ln(S_x - u) + \varepsilon\ln(S_x - \varepsilon u) + \varepsilon^2\ln(S_x - \varepsilon^2 u)$\\
    $w_x=u_y$ &here $\varepsilon = \exp\big(\frac{2\pi i}{3}\big)$  \\ 
      \hline        $h(u,w) =  \sigma(u)e^w$ &  \\ 
 ${\rm System}\ (\ref{eq:pde2}):$ & $S_t = \sigma(u)e^wG_u(S_x, u)$ \\
    $u_t=e^w(\sigma''u_x+2\sigma'u_y+\sigma w_y)$ & $S_y=G(S_x, u)$\\
    $w_x=u_y$ & here $\sigma(u)=\sigma(u;0,g_3)$\\ 
      \hline 
     \end{tabular}
\end{center}
In the last case the function $G(p, u)$ is defined by  the equations
\begin{equation}\label{fl}
G_{p}=\frac{G_{uu}}{G_u}-\zeta(u), \quad G_{uuu}G_u-2G_{uu}^2+2\wp(u) G_u^2=0
\end{equation}
where $\zeta$ and $\wp$ are the Weierstrass functions (equianharmonic case $g_2=0$). The general solution of these equations is given by the formula
\begin{equation}\label{Gpu}
G(p, u)=\ln \sigma (\lambda(p-u))+\epsilon \ln \sigma (\lambda(p-\epsilon u))+\epsilon^2 \ln \sigma (\lambda(p-\epsilon^2 u))
\end{equation}
where $\epsilon=e^{2\pi i/3}=-\frac{1}{2}+i\frac{\sqrt 3}{2}$ and $\lambda=\frac{i}{\sqrt 3}$. Note that the degeneration $g_3\to 0,\ \sigma(u)\to u$ takes the Lax pair corresponding to the Hamiltonian density $h=\sigma(u)e^w$ to the Lax pair for the density $h=ue^w$.
We refer to the Appendix  for a proof that formula (\ref{Gpu}) indeed solves the equations (\ref{fl}): this requires some non-standard identities for equianharmonic elliptic functions.

\subsection{Commuting flows: proof of Theorem \ref{t3}}
\label{sec:comm}
Our aim is to show that every integrable system (\ref{eq:pde2}) possesses a commuting flow of the form
(\ref{com}),
\begin{align}
u_{\tau} &= a(u,w,v)u_x + b(u,w,v)u_y + c(u,w,v)w_y + d(u,w,v)v_y, \nonumber\\  
w_x &= u_y, \nonumber \\
v_x &= (p(u,w))_y. \nonumber
\end{align}
Here $\tau$ is the higher `time' variable and $v=\partial_x^{-1}\partial_y p(u, w)$ is a new nonlocality (to be determined). Due to the presence of nonlocal variables, direct computation of compatibility condition $u_{t\tau}=u_{\tau t}$ is not straightforward. Therefore, we adopt a different approach and require that the combined system (\ref{eq:pde2}) $\cup$ (\ref{com}),
\begin{align}
u_t &= (h_u)_x+(h_w)_y, \label{e1} \\
u_{\tau} &= a(u,w,v)u_x + b(u,w,v)u_y + c(u,w,v)w_y + d(u,w,v)v_y, \label{e2}\\  
w_x &= u_y, \label{e3} \\
v_x &= (p(u,w))_y, \label{e4}
\end{align}
possesses  hydrodynamic reductions. 
Thus, we  seek multiphase solutions of the form  $u = u(R^1,\ldots,R^n)$, $w= w(R^1,\ldots,R^n)$ and $v=v(R^1,\ldots,R^n)$ where the  Riemann invariants $R^i$ satisfy a triple of commuting systems of hydrodynamic type: 
$$
R^i_y = \mu^i(R)R^i_x, \quad R^i_t = \lambda^i(R)R^i_x, \quad R^i_{\tau} = \eta^i(R)R^i_x.
$$
We recall that the commutativity conditions are equivalent to 
\begin{align}
\frac{\partial_j\mu^i}{\mu^j - \mu^i} = \frac{\partial_j\lambda^i}{\lambda^j - \lambda^i}=\frac{\partial_j\eta^i}{\eta^j - \eta^i}. \label{eq:thirdflow}
\end{align}
Following the same procedure as in Section \ref{sec:hydro}, from equations (\ref{e1}) and (\ref{e3})  we  obtain the relations
$\partial_iw=\mu^i\partial_iu$,  the GT-system (\ref{GT1}), (\ref{GT2}), and the integrability conditions (\ref{int}) for the Hamiltonian density $h(u, w)$. Similarly, equation (\ref{e4}) implies
 $$
 \partial_iv = (p_u\mu^i + p_w(\mu^i)^2)\partial_iu,
 $$
and the compatibility condition  $\partial_j \partial_iv=\partial_i  \partial_jv$ results in the relations
$$
h_{uuw}p_w - h_{ww}p_{uu} = 0, \quad
h_{uww}p_w - h_{ww}p_{uw} = 0, \quad
h_{www}p_w - h_{ww}p_{ww} = 0. 
$$
Modulo unessential constants of integration (which can be removed by equivalence transformations) these relations uniquely specify the nonlocality:
\begin{equation*}
p(u,w) = h_w(u,w). 
\end{equation*}   
Finally, equation (\ref{e2}) gives an additional dispersion relation,
$$
\eta^i = a + b\mu^i + (c + p_{u}d)(\mu^i)^2 + p_{w}d(\mu^i)^3.
$$
Substituting $\eta^i$ into the commutativity conditions (\ref{eq:thirdflow}) we obtain the following set of relations:
\begin{align}
p_w^2d_v &= 0, \label{eq:eta1}\\
\big((p_{ww}d + p_wd_w) + p_wp_ud_v\big)h_{ww} &= 2p_w dh_{www}, \label{eq:eta2}\\
(p_{uw}d + p_wd_u)h_{ww} &= 2p_wdh_{uww},\label{eq:eta3}\\
h_{ww}(c_v + p_ud_v)p_w &= p_wdh_{www},\label{eq:eta4} \\ 
h_{ww}\big((c_w + p_{uw}d + p_ud_w) + (c_v +p_ud_v)p_u\big) &= 5h_{uww}dp_w + ch_{www} + p_{u}dh_{www},\label{eq:eta5}\\
h_{ww}b_vp_w &= 2p_wdh_{uww},\label{eq:eta6} \\
h_{ww}(c_u + p_{uu}d + p_ud_u)&= 4p_wdh_{uuw} + ch_{uww} + p_udh_{uww}, \label{eq:eta7}\\
h_{ww}a_vp_w &= p_wdh_{uuw}, \label{eq:eta8} \\
h_{ww}(b_w + b_vp_u) &= 4p_wdh_{uuw} + 2ch_{uww} + 2p_udh_{uww},\label{eq:eta9}\\
h_{ww}b_u &= 2p_wdh_{uuu} + 2ch_{uuw} + 2p_udh_{uuw}, \label{eq:eta10} \\
h_{ww}(a_w + a_vp_u) &= p_wdh_{uuu} + ch_{uuw} + p_udh_{uuw}, \label{eq:eta11} \\
h_{ww}a_u &=ch_{uuu} + p_udh_{uuu}. \label{eq:eta12} 
\end{align}
Using the fact that $p = h_w$ we solve these relations modulo the integrability conditions (\ref{int}), recall that $h_{ww}\ne 0$. Equation (\ref{eq:eta1}) gives $d_v = 0$. 
Equations (\ref{eq:eta2}, \ref{eq:eta3}) imply
\begin{align*}
dh_{www} - h_{ww}d_w &= 0,\;\;\;\;
dh_{uww} - h_{ww}d_u = 0, 
\end{align*}
which  can be solved for $d$:  
\begin{equation*}
d = \delta h_{ww}, 
\end{equation*}
for some constant $\delta$ (which will be set equal to $2$ in what follows). 
Equation (\ref{eq:eta4}) gives 
\begin{equation*}
c_v = d\frac{h_{www}}{h_{ww}}.
\end{equation*}
Setting $c = d\frac{h_{www}}{h_{ww}}v + c_1$ for some  $c_1 = c_1(u,w)$ and substituting into equation (\ref{eq:eta7}) we find 
\begin{equation*}
c_1 = 3dh_{uw} + c_2(w)h_{ww}.
\end{equation*}
Substituting  $c =d\frac{h_{www}}{h_{ww}}v +3dh_{uw} + c_2(w)h_{ww}$  into equation (\ref{eq:eta5}) we find 
\begin{align*}
(c_2)_w &= d\bigg(\frac{h_{uww}h_{ww} - h_{www}h_{uw}}{h_{ww}^2}\bigg)
= d\frac{\partial}{\partial w}\bigg(\frac{h_{uw}}{h_{ww}}\bigg).
\end{align*}
It turns out that modulo the integrability conditions $(c_2)_w$ is a constant. If we set 
\begin{equation*}
\alpha = d\frac{\partial}{\partial w}\bigg(\frac{h_{uw}}{h_{ww}}\bigg),
\end{equation*}
the final formula for $c$ can be written as  
\begin{equation*}
c = d\frac{h_{www}}{h_{ww}}v + 3dh_{uw} + \alpha wh_{ww}. 
\end{equation*}
The equations for the  coefficients $a$ and $b$ cannot be integrated explicitly; rearranging the remaining equations gives the following final result: 
\[
  \left.
  \begin{cases}
    a_u = \frac{h_{uuu}}{h_{ww}}(c + dh_{uw}), &  \\
    a_w = \frac{h_{uuw}}{h_{ww}}c + dh_{uuu}, &  \\
    a_v = d\frac{h_{uuw}}{h_{ww}}. & 
  \end{cases}
  \right.
  \left.
  \begin{cases}
    b_u = 2\big(dh_{uuu} + \frac{h_{uuw}}{h_{ww}}(c + dh_{uw})\big), &  \\
    b_w = 2\big(\frac{h_{uww}}{h_{ww}}c + 2dh_{uuw}\big), &  \\
    b_v = 2d\frac{h_{uww}}{h_{ww}}. & 
  \end{cases}
  \right.
\]
\begin{align*}
 c = d\frac{h_{www}}{h_{ww}}v + 3dh_{uw} + \alpha w h_{ww}, \;\;\; d = \delta h_{ww}, \;\;\; \alpha = d\frac{\partial}{\partial w}\bigg(\frac{h_{uw}}{h_{ww}}\bigg).
\end{align*}
The equations for $a$ and $b$ are consistent  modulo integrability conditions (\ref{int}). This proves the existence of commuting flows (\ref{com}).

\medskip

\noindent{\bf Hamiltonian formulation of commuting flows.} Our next goal in to show that the obtained commuting flow can be cast into Hamiltonian form  
\begin{equation}
u_{\tau} = \partial_x\left(\frac{\delta F}{\delta u}\right), \quad F=\int f(u, w, v)\ dxdy,
\label{HF}
\end{equation}
with the nonlocal variables $w, v$ defined by $w_x=u_y, \ v_x = (h_w)_y$. More precisely, we claim that the commuting density $f$ is given by the formula
$$
f(u, w, v)=vh_w+g(u, w)
$$
where the function $g(u, w)$ is yet to be determined. We have
$$
\frac{\delta F}{\delta u} = 2vh_{uw}+g_u + \partial_x^{-1}\partial_y(2vh_{ww}+g_w),  
$$
so that equation (\ref{HF}) takes the form 
\begin{align}
u_{\tau} &= (2vh_{uw}+g_u)_x + (2vh_{ww}+g_w)_y, \label{eq:hamtau2}\\
w_x &= u_y, \nonumber \\
v_x &= (h_w)_y.\nonumber
\end{align}
Explicitly, (\ref{eq:hamtau2}) gives
\begin{align*}
u_{\tau} &= (2vh_{uuw} + g_{uu})u_x + (4vh_{uww} + 2g_{uw} + 2(h_{uw})^2)u_y \\
&+ (2h_{uw}h_{ww} + 2vh_{www} + g_{ww})w_y + 2h_{ww}v_y.
\end{align*}
Comparing this with (\ref{e2}) we thus require
\begin{align}
a &= 2vh_{uuw} + g_{uu},  \nonumber\\
b &= 4v h_{uww} + 2g_{uw} + 2h_{uw}^2 ,\nonumber\\
c &= 2vh_{www} + 2h_{uw}h_{ww} + g_{ww}, \nonumber\\
d &= 2h_{ww}. \nonumber
\end{align} 
Using the expressions  for $a,b,c,d$ calculated above we obtain the equations for $g(u, w)$:
\begin{align*}
g_{ww} &= 4h_{uw}h_{ww} + \alpha wh_{ww}, \\
g_{uuu} &= 8h_{uw}h_{uuu} + \alpha wh_{uuu}, \\
g_{uuw} &= 6h_{uw}h_{uuw} + 2h_{ww}h_{uuu} + \alpha w h_{uuw};
\end{align*}
note that these equations are consistent modulo the integrability conditions (\ref{int}). This finishes the proof of Theorem \ref{t3}.

\subsection{Commuting flows and dispersionless Lax pairs}
\label{sec:dispcomm}

In this section we calculate  commuting flows of integrable systems (\ref{eq:pde2}) and construct their dispersionless Lax pairs. 

\medskip

\noindent{\bf 1. Hamiltonian density $h(u,w) = \frac{1}{6}u^3 + \frac{1}{2}w^2$.} The commuting density is
 \begin{equation*}
 f(u,w,v) = vw + u^2w.
 \end{equation*}
Commuting flow has the form (note that $\alpha =0$):
\begin{align*}
u_{\tau} &= 2wu_x + 4uu_y+2v_y, \\  
w_x &= u_y,  \\
v_x &= w_y.
\end{align*}
Dispersionless Lax pair:
\begin{align*}
S_y &= \frac{1}{2}S_x^2+u,\\
S_{\tau} &=  \frac{1}{2}S_x^4+ 2uS_x^2+ 2wS_x +2u^2 + 2v.  
\end{align*}

\medskip

\noindent{\bf 2. Hamiltonian density $h(u,w) = w^2 +  u^2w - \frac{1}{4}u^4$.} The commuting density is
 \begin{equation*}
 f(u,w,v) = 2wv +  u^2v + 8 uw^2 -\frac{8}{5}u^5.
 \end{equation*}
Commuting flow has the form (note that $\alpha =0$):
\begin{align*}
u_{\tau} &= (4 v-32u^3)u_x + (32 w+8u^2)u_y+24 uw_y+4v_y, \\  
w_x &= u_y,  \\
v_x &= (2w+ u^2)_y.
\end{align*}
Dispersionless Lax pair:
\begin{align*}
S_y &=  u S_x + \frac{1}{4}S_x^4, \\
S_{\tau} &= (4 v + 32  uw + 16 u^3)S_x + (8 w + 24 w^2)S_x^4 + 8 u S_x^7 + \frac{4}{5}S_x^{10}.
\end{align*}

\medskip

\noindent{\bf 3. Hamiltonian density $h(u,w) = uw^2 +  \beta u^7$.}  The commuting density is
\begin{equation*}
f(u,w,v) = 2uwv + 4uw^3 + 20\beta u^7w. 
\end{equation*}
Commuting flow has the form (note that $\alpha =4$):
\begin{align*}
u_{\tau} &= 840\beta u^5w u_x + (8v+32w^2+280\beta u^6)u_y+32uw w_y+4uv_y, \\  
w_x &= u_y,  \\
v_x &= (2uw)_y.
\end{align*}
Dispersionless Lax pair:
\begin{align*}
S_y &= u^2 \wp(S_x), \\
S_{\tau} &= (8u^2v + 32u^2w^2  + 16u^5w\wp'(S_x) + 8u^8\wp^3(S_x))\wp(S_x),
\end{align*}
where $\wp'^2 = 4\wp^3-35\beta$.

\medskip

\noindent{\bf 4. Hamiltonian density $h(u,w) = e^w$.} The commuting density is
\begin{equation*}
f(u,w,v) = ve^w.
\end{equation*}
Commuting flow has the form (note that $\alpha =0$):
\begin{align*}
u_{\tau} &= 2ve^w w_y+2e^wv_y, \\  
w_x &= u_y,  \\
v_x &= (e^w)_y.
\end{align*}
Dispersionless Lax pair:
\begin{align*}
S_y &=  -\ln(S_x + u),\\
S_{\tau} &= \frac{-2ve^w}{S_x + u} + \frac{e^{2w}}{(S_x + u)^2}.
\end{align*}

\medskip

\noindent{\bf 5. Hamiltonian density $h(u,w)= ue^w$.} The commuting density is
\begin{equation*}
f(u,w,v) = uve^w+ ue^{2w}. 
\end{equation*} 
Commuting flow has the form (note that $\alpha =0$):
\begin{align*}
u_{\tau} &= (4ve^w+6e^{2w})u_y+(2uve^w+6ue^{2w}) w_y+2ue^wv_y, \\  
w_x &= u_y,  \\
v_x &= (ue^w)_y.
\end{align*}
Dispersionless Lax pair:
\begin{align*}
S_y &= \ln(S_x - u) + \varepsilon\ln(S_x - \varepsilon u) + \varepsilon^2\ln(S_x - \varepsilon^2 u), \\
S_{\tau}&= \frac{3u^2e^wS_x(2u^3v - 2vS_x^3 -3e^wS_x^3)}{(S_x^3 - u^3)^2}.
\end{align*}

\medskip

\noindent{\bf 6. Hamiltonian density $h(u,w) = \sigma(u)e^w$.} The commuting density is
\begin{equation*}
f(u,w,v) = v\sigma(u)e^w + \sigma(u)\sigma'(u)e^{2w}.
\end{equation*}
Commuting flow has the form (note that $\alpha =0$):
\begin{align*}
u_{\tau} &= (2v\sigma''e^w+(\sigma \sigma')''e^{2w})u_x+(4v\sigma'e^w+(4\sigma \sigma''+6\sigma'^2)e^{2w})u_y+(2v\sigma e^w+6\sigma \sigma'e^{2w}) w_y+2\sigma e^wv_y, \\  
w_x &= u_y,  \\
v_x &= (\sigma e^w)_y.
\end{align*}
Dispersionless Lax pair:
\begin{align*}
S_y &=G(S_x, u)\\
S_{\tau} &= 2[ve^w\sigma(u) + e^{2w}\sigma(u)\sigma'(u)]G_u(S_x, u) - e^{2w}\sigma(u)^2G_{uu}(S_x, u),
\end{align*}
here $G(S_x, u)$ is defined by equations ({\ref{fl}).

\section{Dispersive deformations} 

Dispersive deformations of hydrodynamic type systems in $1+1$ dimensions were thoroughly investigated in \cite{Dubrovin4, Dubrovin5, Dubrovin6, Dubrovin7} based on deformations of the corresponding hydrodynamic symmetries. In $2+1$ dimensions, an alternative approach based on deformations of hydrodynamic reductions was proposed in \cite{FM, FMN}. 

It still remains a challenging problem  to construct dispersive deformations of all Hamiltonian systems (\ref{eq:pde2}) obtained in this paper. In general,  all three ingredients of the construction may need to be deformed, namely, the Hamiltonian operator $\partial_x$, the Hamiltonian density $h(u, w)$ and the nonlocality $w$. Here we give just two examples.

\medskip

\noindent{\bf Example 1:  dKP equation.}
The Hamiltonian density $h= \frac{1}{2}w^2+\frac{1}{6}u^3$ results in the dKP  equation: 
$$
u_t = uu_x +w_y, \quad w_x=u_y.
$$
It possesses an integrable dispersive deformation
$$
u_t = uu_x +w_y-\epsilon^2u_{xxx}, \quad w_x=u_y,
$$
which is the full KP equation (in this section $\epsilon$ denotes an arbitrary deformation parameter). The KP equation corresponds to the deformed
Hamiltonian  density 
\begin{equation*}
h(u,w) =  \frac{1}{2}w^2  + \frac{1}{6}u^3+\frac{\epsilon^2}{2}u_x^2, 
\end{equation*}
while the Hamiltonian operator $\partial_x$ and the nonlocality $w=\partial_x^{-1}\partial_yu$  stay the same. Indeed, we have
$$
u_t = {\partial_x}\frac{\delta H}{\delta u} 
= {\partial_x}\big(\partial_x^{-1}\partial_yw + \frac{1}{2}u^2 - \epsilon^2u_{xx}\big) \\
= uu_x +w_y-\epsilon^2u_{xxx}.
$$

\medskip

\noindent{\bf Example 2:  Boyer-Finley equation.}
The Hamiltonian density $h=e^w$ results in  the dispersionless Toda (Boyer-Finley) equation:
\begin{align*}
u_t &= e^w w_y,\;\;\;\; w_x = u_y.
\end{align*}
It possesses an integrable dispersive deformation
$$
u_t = \left(\frac{1 - T^{-1}}{\epsilon}\right)e^w, \quad
w_x =\left( \frac{T -1}{\epsilon}\right)u,
$$
which is the full Toda equation. Here $T$ and $T^{-1}$ denote the forward/backward $\epsilon$-shifts in the $y$-direction, so that $\frac{T -1}{\epsilon}$ and $\frac{1 - T^{-1}}{\epsilon}$ are the forward/backward discrete $y$-derivatives. The Toda equation corresponds to the deformed nonlocality 
$w=\partial_x^{-1} \frac{T -1}{\epsilon}u$, while the Hamiltonian operator $\partial_x$ and the Hamiltonian density $h=e^w$ stay the same. Indeed, we have
\begin{align*}
\frac{\delta H}{\delta u} = \partial_x^{-1}\bigg(\frac{1 - T^{-1}}{\epsilon}\bigg)e^w,
\end{align*}
so that 
\begin{align*}
u_t =\partial_x \frac{\delta H}{\delta u}=\bigg(\frac{1 - T^{-1}}{\epsilon}\bigg)e^w,
\end{align*}
as required.

\section{Appendix: dispersionless Lax pair for $h=\sigma(u)e^w$}

Here we prove that expression (\ref{Gpu}),
$$
G(p, u)=\ln \sigma (\lambda(p-u))+\epsilon \ln \sigma (\lambda(p-\epsilon u))+\epsilon^2 \ln \sigma (\lambda(p-\epsilon^2 u)),
$$
where $\epsilon=e^{2\pi i/3}=-\frac{1}{2}+i\frac{\sqrt 3}{2}$ and $\lambda=\frac{i}{\sqrt 3}$,
 solves the equations (\ref{fl}), 
\begin{equation*}
G_{p}=\frac{G_{uu}}{G_u}-\zeta(u), \quad G_{uuu}G_u-2G_{uu}^2+2\wp(u) G_u^2=0.
\end{equation*}
\medskip
In what follows we will use the addition formula
\begin{equation}\label{zeta}
\zeta(u+v)=\zeta(u)+\zeta(v)+\frac{1}{2}\frac{\wp'(u)-\wp'(v)}{\wp(u)-\wp(v)}.
\end{equation}
\medskip
We will also need the following  identity:


\noindent {\bf Proposition 1.} {\it In the equianharmonic case, the Weiesrtrass  functions satisfy the identity}
\begin{equation}\label{iden}
\lambda \frac{\wp'(\lambda u)}{\wp(\lambda u)}+3\lambda \zeta(\lambda u)-\zeta(u)=0, \qquad \lambda=\frac{i}{\sqrt 3}.
\end{equation}

\medskip

\centerline{\bf Proof:}
\medskip
Using the standard expansions
$$
\zeta(z)=\frac{1}{z}-\frac{g_3}{140}z^5-\dots, \qquad \wp(z)=\frac{1}{z^2}+\frac{g_3}{28}z^4+\dots,
$$
one can  show that formula (\ref{iden}) holds to high order in $z$ for the specific parameter value $\lambda=\frac{i}{\sqrt 3}$. Therefore, it is sufficient to establish  the differentiated (by $u$) identity (\ref{iden}), namely,
\begin{equation}\label{idp}
-\lambda^2\wp(\lambda u) +\frac{\lambda^2g_3}{\wp^2(\lambda u)}+\wp(u)=0,
\end{equation}
where we have used $\wp''=6\wp^2$ and $\wp'^2=4\wp^3-g_3$. Explicitly, (\ref{idp})
 reads
$$
\wp(iu/\sqrt3)-\frac{g_3}{\wp^2(iu/\sqrt3)}+3\wp(u)=0.
$$
Setting $u=i\sqrt 3 v$ we obtain
\begin{equation}\label{idp1}
\wp(v)-\frac{g_3}{\wp^2(v)}+3\wp(i\sqrt 3 v)=0.
\end{equation}
Thus, it is sufficient to establish (\ref{idp1}). Formulae of this kind appear in the context of complex multiplication for elliptic curves with extra symmetry. Let us begin with the standard invariance properties of the equianharmonic $\zeta$-function:
$$
\zeta(\epsilon z)=\epsilon^2\zeta(z), \qquad \zeta(\epsilon^2 z)=\epsilon \zeta(z);
$$
here $\epsilon=e^{2\pi i/3}=-\frac{1}{2}+i\frac{\sqrt 3}{2}$ is the cubic root of unity. Setting $z=2v$ this gives
$$
\zeta(-v+i\sqrt 3v)=\epsilon^2\zeta(2v), \qquad \zeta(-v-i\sqrt 3 v)=\epsilon \zeta(2v).
$$
Using the addition formula (\ref{zeta}) one can rewrite these relations in the form
$$
-\zeta(v)+\zeta(i\sqrt 3v)+\frac{1}{2}\frac{-\wp'(v)-\wp'(i\sqrt 3 v)}{\wp(v)-\wp(i\sqrt 3v)}=\epsilon^2\zeta(2v)
$$
and
$$
-\zeta(v)-\zeta(i\sqrt 3v)+\frac{1}{2}\frac{-\wp'(v)+\wp'(i\sqrt 3 v)}{\wp(v)-\wp(i\sqrt 3v)}=\epsilon\zeta(2v),
$$
respectively. Adding there relations together (and keeping in mind that $1+\epsilon+\epsilon^2=0$) we obtain
$$
-2\zeta(v)-\frac{\wp'(v)}{\wp(v)-\wp(i\sqrt 3v)}+\zeta(2v)=0.
$$
Using the duplication formula $\zeta(2v)=2\zeta(v)+\frac{3\wp^2(v)}{\wp'(v)} $ this simplifies to
$$
-\frac{\wp'(v)}{\wp(v)-\wp(i\sqrt 3v)}+\frac{3\wp^2(v)}{\wp'(v)}=0,
$$
which is equivalent to (\ref{idp1}) via $\wp'^2=4\wp^3-g_3$.\qed

\bigskip

\noindent {\bf Proposition 2.} {\it Expression (\ref{Gpu}) solves the equations (\ref{fl}).}

\medskip

\centerline{\bf Proof:}
\medskip

Computation of partial derivatives of $G(p, u)$ gives
$$
G_p=\lambda \zeta (\lambda(p-u))+\lambda \epsilon \zeta (\lambda(p-\epsilon u))+\lambda \epsilon^2 \zeta (\lambda(p-\epsilon^2 u)),
$$
$$
G_u=-\lambda \zeta (\lambda(p-u))-\lambda \epsilon^2  \zeta (\lambda(p-\epsilon u))-\lambda \epsilon \zeta (\lambda(p-\epsilon^2 u)).
$$
Using the addition formula (\ref{zeta}), the identity $1+\epsilon+\epsilon^2=0$, and the invariance
$$
\begin{array}{c}
\zeta(\epsilon z)=\epsilon^2\zeta(z), \quad \zeta(\epsilon^2 z)=\epsilon \zeta(z), \\
\wp(\epsilon z)=\epsilon\wp(z), \quad \wp(\epsilon^2 z)=\epsilon^2 \wp(z), \\
\wp'(\epsilon z)=\wp(z), \quad \wp'(\epsilon^2 z)= \wp(z),
\end{array}
$$
we obtain:
$$
\begin{array}{c}
\frac{1}{\lambda}G_p= \zeta (\lambda(p-u))+ \epsilon \zeta (\lambda(p-\epsilon u))+ \epsilon^2\zeta (\lambda(p-\epsilon^2 u))\\
\ \\
=\zeta(\lambda p)-\zeta(\lambda u)+\frac{1}{2}\frac{\wp'(\lambda p)+\wp'(\lambda u)}{\wp(\lambda p)-\wp(\lambda u)}\\
\ \\
+\epsilon\left(\zeta(\lambda p)-\epsilon^2 \zeta(\lambda u)+\frac{1}{2}\frac{\wp'(\lambda p)+\wp'(\lambda u)}{\wp(\lambda p)-\epsilon \wp(\lambda u)}\right)\\
\ \\
+\epsilon^2\left(\zeta(\lambda p)-\epsilon \zeta(\lambda u)+\frac{1}{2}\frac{\wp'(\lambda p)+\wp'(\lambda u)}{\wp(\lambda p)-\epsilon^2 \wp(\lambda u)}\right)\\
\ \\
=-3\zeta(\lambda u)+\frac{\wp'(\lambda p)+\wp'(\lambda u)}{2}
\left(\frac{1}{\wp(\lambda p)- \wp(\lambda u)}+\frac{\epsilon}{\wp(\lambda p)-\epsilon \wp(\lambda u)}+\frac{\epsilon^2}{\wp(\lambda p)-\epsilon^2 \wp(\lambda u)}
\right)\\
\ \\
=-3\zeta(\lambda u)+\frac{\wp'(\lambda p)+\wp'(\lambda u)}{2}\frac{3\wp^2(\lambda u)}{\wp^3(\lambda p)-\wp^3(\lambda u)}\\
\ \\
=-3\zeta(\lambda u)+\frac{\wp'(\lambda p)+\wp'(\lambda u)}{2}\frac{12\wp^2(\lambda u)}{\wp'^2(\lambda p)-\wp'^2(\lambda u)}\\
\ \\
=-3\zeta(\lambda u)+\frac{6\wp^2(\lambda u)}{\wp'(\lambda p)-\wp'(\lambda u)}.
\end{array}
$$
A similar calculation gives:
$$
\begin{array}{c}
-\frac{1}{\lambda}G_u= \zeta (\lambda(p-u))+ \epsilon^2 \zeta (\lambda(p-\epsilon u))+ \epsilon\zeta (\lambda(p-\epsilon^2 u))\\
\ \\
=\zeta(\lambda p)-\zeta(\lambda u)+\frac{1}{2}\frac{\wp'(\lambda p)+\wp'(\lambda u)}{\wp(\lambda p)-\wp(\lambda u)}\\
\ \\
+\epsilon^2\left(\zeta(\lambda p)-\epsilon^2 \zeta(\lambda u)+\frac{1}{2}\frac{\wp'(\lambda p)+\wp'(\lambda u)}{\wp(\lambda p)-\epsilon \wp(\lambda u)}\right)\\
\ \\
+\epsilon\left(\zeta(\lambda p)-\epsilon \zeta(\lambda u)+\frac{1}{2}\frac{\wp'(\lambda p)+\wp'(\lambda u)}{\wp(\lambda p)-\epsilon^2 \wp(\lambda u)}\right)\\
\ \\
=\frac{\wp'(\lambda p)+\wp'(\lambda u)}{2}
\left(\frac{1}{\wp(\lambda p)- \wp(\lambda u)}+\frac{\epsilon^2}{\wp(\lambda p)-\epsilon \wp(\lambda u)}+\frac{\epsilon}{\wp(\lambda p)-\epsilon^2 \wp(\lambda u)}
\right)\\
\ \\
=\frac{\wp'(\lambda p)+\wp'(\lambda u)}{2}\frac{3\wp(\lambda p)\wp(\lambda u)}{\wp^3(\lambda p)-\wp^3(\lambda u)}\\
\ \\
=\frac{\wp'(\lambda p)+\wp'(\lambda u)}{2}\frac{12\wp(\lambda p)\wp(\lambda u)}{\wp'^2(\lambda p)-\wp'^2(\lambda u)}\\
\ \\
=\frac{6\wp(\lambda p)\wp(\lambda u)}{\wp'(\lambda p)-\wp'(\lambda u)}.
\end{array}
$$
To summarise, we have:
$$
G_p=-3\lambda \zeta(\lambda u)+\frac{6\lambda \wp^2(\lambda u)}{\wp'(\lambda p)-\wp'(\lambda u)}, \qquad
G_u=-\frac{6\lambda \wp(\lambda p)\wp(\lambda u)}{\wp'(\lambda p)-\wp'(\lambda u)}.
$$
This gives
\begin{equation}\label{Guu}
\frac{G_{uu}}{G_u}=(\ln G_u)_u=\lambda\frac{\wp'(\lambda u)}{\wp(\lambda u)}+\frac{6\lambda \wp^2(\lambda u)}{\wp'(\lambda p)-\wp'(\lambda u)},
\end{equation}
and the first equation (\ref{fl}), $G_p=\frac{G_{uu}}{G_u}-\zeta(u)$, is satisfied identically due to 
(\ref{iden}). Finally, the second equation (\ref{fl}), $G_{uuu}G_u-2G_{uu}^2+2\wp(u) G_u^2=0$,
which can be written in the equivalent form
$$
-\left(\frac{G_{uu}}{G_u}\right)_u+\left(\frac{G_{uu}}{G_u}\right)^2=2\wp(u),
$$
is satisfied identically due to (\ref{Guu}) and (\ref{idp}). \qed

\section*{Acknowledgements}

We thank Yurii Brezhnev and Maxim Pavlov for clarifying discussions. The research of EVF was supported by the EPSRC grant EP/N031369/1. The research of VSN was supported by the EPSRC grant EP/V050451/1.

\end{document}